\documentclass[conference]{IEEEtran}
\IEEEoverridecommandlockouts
\usepackage{cite}
\usepackage{amsmath,amssymb,amsfonts}
\usepackage{algorithmic}
\usepackage{graphicx}
\usepackage{textcomp}
\usepackage{xcolor}
\usepackage{subcaption}
\usepackage{hyperref}

\def\BibTeX{{\rm B\kern-.05em{\sc i\kern-.025em b}\kern-.08em
    T\kern-.1667em\lower.7ex\hbox{E}\kern-.125emX}}
    
\usepackage{fancyhdr}
\begin{document}

\title{Towards Emotion-Aware Agents For Negotiation Dialogues
\thanks{Research was sponsored by the National Science Foundation and Army Research Office under Cooperative Agreement W911NF-20-2-0053. The views and conclusions contained in this document are those of the authors and should not be interpreted as representing the official policies, either expressed or implied, of the Army Research Office or the U.S. Government. The U.S. Government is authorized to reproduce and distribute reprints for Government purposes notwithstanding any copyright notation herein.}
}

\author{\IEEEauthorblockN{Kushal Chawla$^{1}$\hspace{0.3cm}Rene Clever$^{2}$ \hspace{0.3cm} Jaysa Ramirez$^{3}$\hspace{0.3cm} Gale Lucas$^1$\hspace{0.3cm} Jonathan Gratch$^1$}
\IEEEauthorblockA{\textit{$^{1}$University of Southern California, Los Angeles, USA}\\\textit{$^2$CUNY Lehman College, Bronx, USA}\hspace{0.3cm}\textit{$^3$Rollins College, Winter Park, USA}\\
$^1$\{chawla, lucas, gratch\}@ict.usc.edu\hspace{0.3cm} $^2$rene.clever@lc.cuny.edu\hspace{0.3cm}  $^3$jramirez@rollins.edu}
}

\maketitle
\thispagestyle{fancy}

\begin{abstract}
Negotiation is a complex social interaction that encapsulates emotional encounters in human decision-making. Virtual agents that can negotiate with humans are useful in pedagogy and conversational AI. To advance the development of such agents, we explore the prediction of two important subjective goals in a negotiation -- outcome satisfaction and partner perception. Specifically, we analyze the extent to which emotion attributes extracted from the negotiation help in the prediction, above and beyond the individual difference variables. We focus on a recent dataset in chat-based negotiations, grounded in a realistic camping scenario. We study three degrees of emotion dimensions -- \textit{emoticons}, \textit{lexical}, and \textit{contextual} by leveraging affective lexicons and a state-of-the-art deep learning architecture. Our insights will be helpful in designing adaptive negotiation agents that interact through realistic communication interfaces.
\end{abstract}

\section{Introduction}

Negotiation is a core task for studying emotional feelings and expressions in human decision-making~\cite{gratch2015negotiation}. Being a mixed-motive task, it creates both interpersonal and intrapersonal conflicts for the negotiators. Motivational tensions often arise when negotiations pit aspirations for individual accomplishments against the demands of sustaining social connections. This leaves difficult decisions for the negotiators about working towards their own self-oriented outcomes or making sacrifices for others. Such situations can be fraught with emotional encounters. For instance, as a negotiator strives to get as much as possible for themselves, in doing so, they need their partner to go along as well. However, the partner's willingness to go along is essentially an emotional decision. A negotiator that tries to take too much can annoy their partner, and in turn, hurt their likeability in the eyes of their partners and also the partner's affective evaluation of the outcome (that is, their satisfaction). Instead, it is desirable for the negotiator to strive for maximum performance while ensuring that the partner is satisfied~\cite{oliver1994outcome} and leaves with a positive perception of the partner~\cite{mell2019likeability, aydougan2020challenges}. Therefore, predicting the partner's satisfaction and perception in advance can be crucial for an AI assistant that aims to negotiate with its users. These virtual negotiation agents find broad applications in advancing conversational AI such as the Google Duplex agent that books a haircut appointment over the phone~\cite{leviathan2018google}. Further, such agents can also be useful in increasing access to social skills training~\cite{johnson2019intelligent}. 


A number of prior studies rely on user individual difference attributes to explain negotiation behaviors and outcomes. For instance, this includes studies based on gender~\cite{stuhlmacher1999gender} and personality traits such as Social Value Orientation and Machiavellianism~\cite{xu2017towards}. This research is crucial from the perspective of practical negotiation agents that interact with users on social media platforms since research suggests that demographic and personality attributes can be inferred from past user interactions~\cite{dong2014inferring, ortigosa2011inferring, adali2014predicting}. Thus, they can be used to adapt the behavior of the deployed automated systems. However, relying only on these variables misses out on all the other available information such as affective attributes in the negotiation itself, which may further help in predicting the outcomes. While one might expect that such affect variables would help merely because they are manifestations of the individual differences, it is instead possible that affective factors reflect the recent interaction with the partner and might predict in their own right. 
This leads us to our key research question in this work: \textit{For predicting a negotiator's satisfaction and liking for their partners, is there value in leveraging affective attributes from the negotiation, above and beyond the individual difference variables?}

Most prior efforts in negotiations focus on a menu-driven communication between the human and the agent. Such interfaces allow button clicks for sharing offers or preferences, which helps to keep the design tractable. This concreteness, however, comes at a cost -- it hinders the analysis of several realistic aspects in a negotiation such as persuasion and free-form emotion expression. Due to the inherent design restrictions, emotion expression merely reduces to the use of emoticons and predefined sentence templates~\cite{mell2016iago}. Instead, recent efforts have explored a fundamentally different, chat-based interaction that allows human-agent communication in free-form natural language such as English~\cite{he2018decoupling, chawla2021casino}, having the potential to fill the gap between existing emotion research and the challenges in real-world negotiations.

To this end, we analyze a large-scale, linguistically rich dataset~\cite{chawla2021casino} that is grounded in a realistic campsite negotiation task. The associated metadata about the participants enables us to study individual differences and linguistic dialogue behaviors in the same setup. Going beyond the capabilities of restrictive menu-driven interfaces, we develop three degrees of emotion recognition techniques: \textit{emoticons}, \textit{lexical}, and \textit{contextual} emotion by leveraging recent advancements in deep learning (Section \ref{sec:dataset}). Using correlational and step-wise regression analysis, we quantify the extent to which affective attributes help to account for the variance in the outcomes, beyond just the individual user attributes (Section \ref{sec:results}). We further discuss how our analysis can guide the development of automated negotiation agents and briefly summarise the ethical considerations around such systems (Section \ref{sec:discussion}).

\section{Related Work}
How humans negotiate has been extensively studied in multiple disciplines: Game Theory for identifying optimal behaviors~\cite{nash1950bargaining}, Psychology for understanding human decision-making~\cite{carnevale1992negotiation}, and Computer Science for building automated negotiation agents~\cite{beam1997automated, baarslag2016survey}. Demographics, personality and affective behaviors have played a central role in improving our understanding around negotiations~\cite{stuhlmacher1999gender, xu2017towards} and similar decision-making social interactions~\cite{stratou2015emotional, de2011effect}. For instance, de Melo and colleagues studied human-agent prisoner's dilemma interaction and found that the display of guilt after exploitation can elicit more cooperative responses as compared to an agent that smiled~\cite{de2011effect}.

Both satisfaction and the relationship with the negotiation partner are seen as crucial metrics of negotiation performance. Oliver and colleagues studied how profit expectations relate to satisfaction after the bargaining in buyer-seller interactions, finding that higher expectations had the effect of decreased satisfaction~\cite{oliver1994outcome}. Maintaining a positive relationship with the partner is especially crucial in repeated interactions, where poor relations in earlier negotiations can adversely impact the results of future ones~\cite{aydougan2020challenges}. Relationship also manifests in the context of rapport building~\cite{nadler2003rapport}, favor exchange~\cite{mell2015effective}, and reputation effects~\cite{zacharia2000trust}.

However, most efforts in human-agent negotiations are based on highly structured menu-driven designs such as the IAGO negotiation platform~\cite{mell2016iago}. In many cases, the negotiation is seen as simply an exchange of offers~\cite{baarslag2013evaluating}. Towards more realistic modes of communication, limited efforts have studied chat-based dialogue systems, that are fundamentally different than menu-driven systems and hence, must be studied separately. Initial datasets in this direction looked at game settings~\cite{lewis2017deal}, which allow free-form language use but the lack of any semantic context in such negotiations inhibits any rich personal conversations. He and colleagues focused on more realistic buyer-seller price negotiations for the product listings on Craigslist~\cite{he2018decoupling}. Unfortunately, none of these datasets incorporated the demographics or the personality of the users. Further, all of them looked only at the objective outcomes of the negotiation performance such as the points scored or the final agreed price. Our focus is on the recently released \textit{CaSiNo} dataset that enables linguistically rich conversations in a constrained, tractable environment~\cite{chawla2021casino}. The dataset further contains the metadata of the participants, that we use to develop our measures in this work (Section \ref{sec:dataset}).

Finally, we note that researchers have also looked at face-to-face negotiations~\cite{devault2015toward}, including speech and embodied agents that can be natural extensions to our current analysis in chat-based negotiations.

\section{Dataset and Methods}
\label{sec:dataset}

\begin{figure*}[htbp]
\centering
\begin{subfigure}[b]{0.8\linewidth}
 \centering
\includegraphics[width=\linewidth]{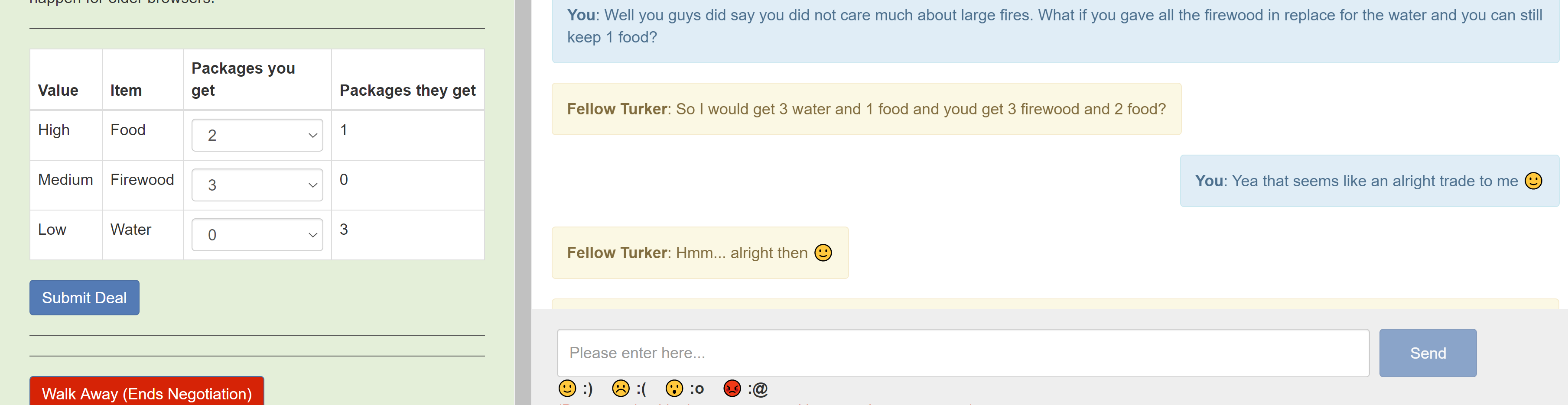}
\end{subfigure}
\caption{Screenshot from the CaSiNo data collection interface. Taken from \cite{chawla2021casino}.}
\label{fig:interface}
\end{figure*}

Our analysis is based on the recently released CaSiNo dataset~\cite{chawla2021casino}. The design is an instance of Multi-Issue Bargaining Task (MIBT), which is a popular model for studying negotiations computationally~\cite{fershtman1990importance,mell2016iago}. MIBT provides a concrete structure to the negotiation and thus, it helps to keep the conversations focused. Further, the task is grounded in a real-world campsite scenario that promotes rich personal conversations in free-form natural language. Unlike prior datasets in chat-based negotiation systems~\cite{asher2016discourse,lewis2017deal,he2018decoupling}, this design enables the study of a number of realistic aspects in negotiations such as persuasion and emotion expression in a constrained, tractable environment.


\subsection{Participants}
\label{sec:participants}
A total of $846$ unique participants were recruited on Amazon Mechanical Turk for collecting the CaSiNo dataset. The participant pool was restricted to the United States. They were paid for their time and also with a performance based bonus. Among the participants, $472$ identified themselves as \textit{Female}, $372$ were \textit{Male}, and \textit{2} belonged to \textit{Other} category. $73.9\%$ of the participants were \textit{White American}, $1.5\%$ were \textit{Native or Islander}, $8.7\%$ were \textit{Asian American}, $8.6\%$ were \textit{Black or African American}, $4.1\%$ were of \textit{Hispanic or Latino} origin, and remaining $3.1\%$ belonged to \textit{Other} category.

\subsection{Design and Procedure}
The CaSiNo dataset contains a total of $1030$ negotiation dialogues in English. In each dialogue, two participants take the role of campsite neighbors and negotiate for additional packages of camping essential items. Following the MIBT structure, there are three issues: \textit{Food}, \textit{Water}, and \textit{Firewood}. Each issue has an available quantity of three, thereby resulting in four \textit{levels} from $0$ to $3$. Before the negotiation begins, each participant is randomly assigned a priority order for the three issues, which is a permutation of \{\textit{High}, \textit{Medium}, \textit{Low}\}. Once the priorities are assigned, the participants also come up with their own justifications for needing or not needing a particular issue such as needing more water for a planned hike or food for the kids. This exercise \textit{prepares} the participants for their upcoming negotiation, and subsequently results in the collection of linguistically rich conversations.

The dialogues in CaSiNo contain $11.6$ utterances with $22$ tokens per utterance, on average. The participants were also allowed to use emoticons for four basic emotions: \textit{Joy}, \textit{Sadness}, \textit{Anger}, and \textit{Surprise}. We provide a screenshot from the chat interface in Figure \ref{fig:interface}. Once the negotiation is over, the researchers record the performance of the participants based on the number of items that they are able to negotiate for. Alongside, the participants are also asked to report their satisfaction with the outcome, and their perception of the negotiation partner. Analyzing these subjective measures of negotiation performance is the primary focus of this work. The dataset further contains the demographics and personality traits for the participants, which are collected prior to the negotiation. We refer the readers to the original CaSiNo paper~\cite{chawla2021casino} for more information on the data collection procedure and a sample dialogue from the dataset.

\subsection{Measures}
We cluster the variables into three categories: 1) \textit{Individual difference}, which captures the demographics and personality of the participants, 2) \textit{Affect Variables}, which comprises all the affective attributes extracted from the negotiation utterances, and 3) \textit{Negotiation Outcomes}, where we describe the two primary dependent variables in our analysis.

\subsubsection{Individual Difference} These variables are based on the responses reported by the participants in a survey before their negotiation. The participants self-identified their demographic attributes while the personality traits are based on the standard tests from the psychology literature.

\textbf{Demographics}: There are two continuous variables: \textit{Age}\footnote{One participant reported the age of $3$, which we believed to be in error and was removed from all our analysis that uses \textit{Age}.} and \textit{Education}. We encoded \textit{Education} as a continuous variable, leveraging the inherent order in the highest level of education. It takes a value from $0$ to $8$, with an increasing level of education.

Further, there are two discrete demographic variables: \textit{Gender} and \textit{Ethnicity}. For our regression analysis, we dummy-coded these variables based on the categories discussed earlier.

\textbf{Personality}: There are two available measures of individual personality differences: Social Value Orientation~\cite{van1997development} and the Big-$5$ personality traits~\cite{goldberg1990alternative}. These personality attributes have been extensively studied in the context of negotiation research~\cite{bogaert2008social, curtis2015relationship}. The Social Value Orientation (SVO) is defined as the stable preferences for certain patterns of outcomes for oneself or others~\cite{mcclintock1978social}. A participant can either be categorized as \textit{Prosocial}, that tend to be cooperative in their interactions, or \textit{Proself}, that tend to serve their individual interests. In the CaSiNo data collection study, SVO was computed using the Triple Dominance Measure~\cite{van1997development}. $463$ participants were classified as \textit{Prosocial}, $364$ as \textit{Proself}, while $19$ were \textit{Unclassified}. The Big-$5$ test consists of five personality dimensions: \textit{Extraversion}, \textit{Agreeableness}, \textit{Conscientiousness}, \textit{Emotional Stability}, and \textit{Openness to Experiences}. These dimensions were computed using the Ten-Item Personality Inventory~\cite{ehrhart2009testing}. Each dimension takes a value between $1$ and $7$.

\subsubsection{Capturing Affect}
Natural language based negotiations provide exciting avenues for Affective Computing research. Unlike menu-driven systems, where communication is highly restricted, language allows free expression of emotion and other affective attributes. In order to capture these attributes in CaSiNo, we consider three different degrees of affect recognition techniques: emoticon counts, use of emotive vocabulary based on affect lexicons (specifically LIWC~\cite{pennebaker2001linguistic}), and utterance-level emotion expressions (based on a pretrained deep learning model called T5~\cite{raffel2020exploring}). We discuss these approaches below. We will later compare these methods through our regression analysis for predicting satisfaction and likeness.

\textbf{Emoticons}: Emoticons provide a structured way for emotion expression. Due to this reason, emoticons have been heavily used in menu-driven negotiation research. For instance, the IAGO platform for human-agent negotiations allows human participants to use the emoticons via button clicks~\cite{mell2016iago}. For CaSiNo, the participants were allowed to use shorthands for typing emoticons directly in the chat interface for four basic emotions: \textit{Joy}, \textit{Sadness}, \textit{Anger}, and \textit{Surprise} (see Figure \ref{fig:interface}). For a given participant and the negotiation dialogue, we count the number of emoticons used by the participant for each of these types and use that as a continuous measure of emotion expression. Approximately $15\%$ of the utterances in CaSiNo make use of one or more emoticons. Among those that use it, nearly $80\%$ use \textit{Joy} emoticon.

Based on a manual inspection of the utterances, we find that the participants tend to use \textit{Joy} emoticon in a number of scenarios, including small talk at the beginning or end of the conversation when showing agreement, and in some cases, interestingly, also when rolling out a strict offer. \textit{Sadness} tends to be used when showing disagreement, showing empathy, or emphasizing personal needs. Some cases where the participants expressed \textit{Surprise} are when pointing out strange behavior from their partner or showing empathy towards a specific need of their partner. \textit{Anger} is used in the cases of strong disagreement where for instance, the partner does not empathize with the personal requirements of the participants or when they receive a very unfair offer that largely benefits the negotiation partner.

\textbf{LIWC}:
We now go beyond the capabilities of menu-driven systems by extracting emotion attributes from the text in the utterances of the participants. Specifically, we look at word-level expressions by leveraging affect lexicons from the literature. Specifically, we make use of the Linguistic Inquiry and Word Count lexicon, popularly referred to as LIWC in the literature~\cite{pennebaker2001linguistic}. LIWC consists of a number of word vocabularies corresponding to everyday use that reflect our feelings and personality. Given our focus on affective attributes, we use four LIWC dimensions in this work: \textit{Positive Emotions}, \textit{Sadness}, \textit{Anger}, and \textit{Anxiety}. To compute the corresponding scores for a participant in a given dialogue, we simply count the number of words in the utterances of the participant that are present in the corresponding LIWC lexicons.

As one might expect, we find a number of instances where the participants express emotions without explicitly making use of emoticons. For instance, an excerpt where the participant clearly expresses anger without using the \textit{anger} emoticon: `Are you mad. with out water what we do...'. Hence, with LIWC lexicons, we are able to capture more emotive utterances than by simply looking at the emoticon usage. Nearly $60\%$ of the utterances make use of at least one emotive word from our four LIWC categories. 

Since LIWC variables are based on word usage, they provide a highly interpretable way of extracting emotions directly from the text. However, merely relying on word-level traits can still miss out on phrase-level or utterance-level emotion expressions. Hence, we next discuss how we make use of a state-of-the-art deep learning model from Natural Language Processing (NLP) to identify emotion in CaSiNo dialogues.

\textbf{Utterance-level emotion}:
Large deep learning models based on the highly parallelizable Transformer architecture have gained enormous popularity in the field of NLP in recent years~\cite{vaswani2017attention}. These models have achieved huge success on a variety of tasks ranging from sentiment analysis to machine translation, fundamentally changing the way researchers approach these tasks. The primary way of training a model for a specific task consists of two steps: 1) \textit{Pre-train} a large language model with freely available text data on the web in an unsupervised or a self-supervised manner, and then 2) \textit{Fine-tune} the model on the task-specific dataset based on supervised learning.

Leveraging these recent advancements, we make use of the Text-to-Text Transfer Transformer (T5) model~\cite{raffel2020exploring}. T5 is an encoder-decoder architecture that is pretrained on a variety of NLP tasks by converting all the tasks in the same text-to-text format. No matter the task, the input to the model is a text sequence along with a task identifier prefix, and the target output is also a text sequence. Classification tasks can also be converted in this way by simply using a target output sequence that consists of a single word corresponding to the ground-truth label. This approach swiftly unifies a number of tasks, paving the way for parameter sharing throughout the model architecture. T5 achieves the state-of-the-art performance on a number of NLP tasks such as classification, text summarization, and question answering.

For our purpose, we first pre-process the utterances in CaSiNo by removing all the emoticons. We then use the publicly available T5 model\footnote{https://huggingface.co/mrm8488/t5-base-finetuned-emotion} for making predictions. This model is \textit{fine-tuned} on an emotion recognition dataset based on social interactions on Twitter~\cite{saravia2018carer}. Given an input text sequence, the model classifies it into one of the six classes: \textit{Joy}, \textit{Love}, \textit{Sadness}, \textit{Fear}, \textit{Anger}, and \textit{Surprise}. The model performs well on the test set of the emotion data, achieving an accuracy of $93\%$ and a macro F1 score of $90\%$.

The output prediction of the model is based on the confidence scores that it generates for all the emotion labels, such that the label with the highest confidence is chosen as the prediction. In our case, since we are primarily interested in extracting the emotion scores for the utterances, we do not rely on the final predictions of the model. Instead, we directly use these confidence scores (probability values between $0$ and $1$) as the corresponding scores for each label. Given a negotiation dialogue and a participant, we simply sum up these confidence scores for all the participant's utterances and use these as the continuous measures for the six emotion dimensions. We refer to these dimensions (and interchangeably, this method) as \textit{T5-Emotion} in this paper.

All the three methods described above capture the emotion expression in CaSiNo dataset, in fundamentally very different ways. To gain further insight into these methods, in Section \ref{sec:analysis}, we will analyze the relationship between them through a combination of quantitative and qualitative techniques. First, we complete our description for measures by elaborating on the subjective negotiation outcomes.

\begin{table*}[thbp]
\caption{Correlations of \textit{T5-Emotion} variables (a) among each other, and (b) with corresponding \textit{Emoticon} and \textit{LIWC} variables. We only consider the dimensions that are common across all methods. * denotes $p<0.05$. ** denotes $p<0.01$.}
\begin{center}
\begin{subtable}{0.3\linewidth}
\caption{}
\centering
\scalebox{0.8}{
\begin{tabular}{|l|c|c|c|}
\hline
& \textbf{Joy} & \textbf{Sadness} & \textbf{Anger}\\ \hline
\textbf{Joy} & $1$ & $-.288$** & $-.295$** \\ 
\textbf{Sadness} & $-.288$** & $1$ & $.277$**\\ 
\textbf{Anger} & $-.295$** & $.277$** & 1\\ \hline
\end{tabular}}
\end{subtable}
\begin{subtable}{0.6\linewidth}
\caption{}
\centering
\scalebox{0.8}{
\begin{tabular}{|l|c|c|c|c|c|c|}
\hline
\textbf{T5-Emotion} & \multicolumn{3}{|c|}{\textbf{Emoticon Usage}} & \multicolumn{3}{|c|}{\textbf{LIWC Variables}} \\ \hline
& \textbf{Joy} & \textbf{Sadness} & \textbf{Anger} & \textbf{Positive Emotions} & \textbf{Sadness} & \textbf{Anger} \\
\textbf{Joy} & $.062$** & $-.043$ & $-.040$ & $.387$** & $-.125$** & $-.106$** \\ 
\textbf{Sadness} & $-.010$ & $.213$** & $.086$** & $-.033$ & $.550$** & $.033$\\ 
\textbf{Anger} & $-.052$* & $.113$** & $.111$** & $-.185$** & $.114$** & $.169$**\\ \hline
\end{tabular}}
\end{subtable}
\label{tab:correlations-among-emotion}
\end{center}
\end{table*}

\begin{table*}[htbp]
\caption{Top 5 words based on log-odds ratio for (a) Emoticon (b) LIWC and (c) T5-Emotion methods. * denotes $p<0.05$. ** denotes $p<0.01$. *** denotes $p<0.001$.}
\begin{center}
\begin{subtable}{\linewidth}
\caption{}
\centering
\scalebox{0.8}{
\begin{tabular}{|c|c|c|c|c|c|c|c|}
\hline
\multicolumn{2}{|c|}{\textbf{Joy}} & \multicolumn{2}{|c|}{\textbf{Sadness}} & \multicolumn{2}{|c|}{\textbf{Anger}} & \multicolumn{2}{|c|}{\textbf{Surprise}}\\
\textbf{Word} & \textbf{Z-score} & \textbf{Word} & \textbf{Z-score} &\textbf{Word} & \textbf{Z-score} &\textbf{Word} & \textbf{Z-score} \\ \hline 
hope&$2.89$**&sorry & $2.10$*&animal & $1.36$&drive & $1.43$ \\
buddy&$2.72$**&spilled & $1.84$&unacceptable & $0.98$&bargain & $1.43$ \\
awesome&$2.71$**&n't & $1.82$&covid-$19$ & $0.84$&awful & $1.32$ \\
great&$2.60$**&reconsider & $1.38$&word & $0.84$&function & $1.32$ \\
thank&$2.53$*&stopped & $1.30$&joking & $0.84$&snack & $1.23$ \\
\hline
\end{tabular}}
\end{subtable}
\begin{subtable}{\linewidth}
\caption{}
\centering
\scalebox{0.8}{
\begin{tabular}{|c|c|c|c|c|c|c|c|}
\hline
\multicolumn{2}{|c|}{\textbf{Positive Emotions}} & \multicolumn{2}{|c|}{\textbf{Sadness}} & \multicolumn{2}{|c|}{\textbf{Anger}} & \multicolumn{2}{|c|}{\textbf{Anxiety}}\\
\textbf{Word} & \textbf{Z-score} & \textbf{Word} & \textbf{Z-score} &\textbf{Word} & \textbf{Z-score} &\textbf{Word} & \textbf{Z-score} \\ \hline 
good & $11.89$*** &sorry&$9.93$***&hate &$2.08$*&worried &$7.18$*** \\
well & $11.87$*** &low&$6.16$***& greedy&$1.58$ &afraid& $4.07$***\\
okay & $9.90$*** &alone& $4.43$***&critical&$1.35$&worry & $3.5$***\\
great& $9.86$*** &sugar& $2.44$*&tricky&$1.19$&risk & $3.22$**\\
ok& $9.68$*** &hear &$2.43$*&bother &$1.06$&confused & $2.54$*\\ \hline
\end{tabular}}
\end{subtable}
\begin{subtable}{\linewidth}
\caption{}
\centering
\scalebox{0.8}{
\begin{tabular}{|c|c|c|c|c|c|c|c|c|c|c|c|}
\hline
\multicolumn{2}{|c|}{\textbf{Joy}} & \multicolumn{2}{|c|}{\textbf{Love}} & \multicolumn{2}{|c|}{\textbf{Sadness}} & \multicolumn{2}{|c|}{\textbf{Fear}} &
\multicolumn{2}{|c|}{\textbf{Anger}} & \multicolumn{2}{|c|}{\textbf{Surprise}}\\
\textbf{Word} & \textbf{Z-score} & \textbf{Word} & \textbf{Z-score} &\textbf{Word} & \textbf{Z-score} &\textbf{Word} & \textbf{Z-score} & \textbf{Word} & \textbf{Z-score} & \textbf{Word} & \textbf{Z-score} \\ \hline 
you & $8.62$***&hot & $6.4$***&sorry & $10.63$***&worried & $8.72$***&cold & $11.5$***&funny & $1.93$ \\
good & $7.68$***&generous & $3.04$**&unfortunately & $4.68$***&afraid & $7.05$***&no & $5.76$***&interesting & $1.89$ \\
great & $7.33$***&hotter & $1.89$&dehydrated & $4.03$***&concerned & $4.48$***&need & $4.77$***&surprise & $1.08$ \\
hello & $6.84$***&lovely & $1.56$&low & $3.12$***&dark & $3.88$***&thirsty & $4.63$***&ha & $1.03$ \\
sounds & $6.81$***&liking & $1.34$&suffer & $2.98$**&scared & $3.58$***&hungry & $4.36$***&status & $0.85$ \\
\hline
\end{tabular}}
\end{subtable}
\label{tab:liwc-log-odds}
\end{center}
\end{table*}

\begin{table}[htbp]
\caption{High confidence sample predictions for \textit{T5-Emotion} model that went \textit{undetected} for both \textit{Emoticon} and \textit{LIWC} methods.}
\begin{center}
\scalebox{0.8}{
\begin{tabular}{|l|p{6cm}|}
\hline
\textbf{Prediction} & \textbf{Sample Utterances} \\ \hline
\textbf{Joy} &1) I think that sounds reasonable to me.\\
&2) I can make that deal work.\\
& 3) You sound very resourceful!  Sounds like we have a deal then, yes? \\ \hline
\textbf{Sadness} &1) I feel like this deal just keeps getting worse for me. I won't take less than 2 waters.\\
&2) That is unfortunate. How about I give you all my firewood for all your food?\\
& 3) Oh, that would make things difficult! \\ \hline
\textbf{Anger} & 1) You are not getting everything, thats just selfish\\
&2) I am about to walk, you are being so unfair\\
& 3) You gave me one on each, thats unfair man!\\ \hline
\end{tabular}}
\label{tab:t5-sample-predictions}
\end{center}
\end{table}

\subsubsection{Negotiation Outcomes}
We focus on two outcome variables: 1) Satisfaction (How satisfied are you with the negotiation outcome?), and 2) Likeness (How much do you like your opponent?). Both variables were self-reported by the participants after the negotiation, using a 5-point Likert scale. We coded the responses from $1$ to $5$ to create the continuous measures. We present the means and standard deviations for all the continuous variables in Table \ref{tab:variables-stats}.

\begin{table}[htbp]
\caption{Statistics and correlations with outcome variables for all continuous measures. * denotes $p<0.05$. ** denotes $p<0.01$.}
\begin{center}
\scalebox{0.9}{
\begin{tabular}{|l|c|c|c|c|}
\hline
\textbf{Variable} & \textbf{Mean} & \textbf{Std.} & \multicolumn{2}{|c|}{\textbf{Correlations}}\\
 & & & \textbf{Satisfaction} & \textbf{Likeness}\\
\hline
\multicolumn{5}{|c|}{\textbf{Individual Difference}} \\
\textbf{Age} & $36.97$ & $10.81$ & $.110$** & $.158$**\\
\textbf{Education} & $5.23$ & $1.67$ & $-.004$ & $-.009$\\
\textbf{Extraversion} & $3.69$ & $1.70$ & $.055$* & $.060$**  \\
\textbf{Agreebleness} & $5.27$ & $1.29$ & $.067$** & $.099$** \\
\textbf{Conscientiousness} & $5.60$ & $1.26$ & $.026$ & $.046$*\\
\textbf{Emotional Stability} & $4.91$ & $1.56$ & $.062$** & $.087$**\\
\textbf{Openness to Experiences} & $5.04$ & $1.31$ & $.027$ & $.042$\\ \hline

\multicolumn{5}{|c|}{\textbf{Emoticons}} \\

\textbf{Joy} & $.77$ & $1.24$ & $.040$ & $.075$**\\
\textbf{Sadness} & $.13$ & $.42$ & $-.151$** & $-.170$**\\
\textbf{Anger} & $.02$ & $.18$ & $-.074$** & $-.108$**\\
\textbf{Surprise} & $.07$ & $.27$ & $-.062$** & $-.059$**\\ \hline

\multicolumn{5}{|c|}{\textbf{LIWC}} \\

\textbf{Positive Emotions} & $5.58$ & $3.36$ & $.033$ & $.061$** \\ 
\textbf{Sadness} & $.22$ & $.52$ & $-.057$** & $-.091$**\\ 
\textbf{Anger} & $.05$ & $.25$ & $-.047$* & $-.043$\\ 
\textbf{Anxiety} & $.18$ & $.52$ & $-.040$ & $-.014$ \\ \hline

\multicolumn{5}{|c|}{\textbf{T5-Emotion}} \\

\textbf{Joy} & $3.89$ & $1.09$ & $.010$ & $.026$\\ 
\textbf{Love} & $.22$ & $.21$ & $-.010$ & $-.030$\\ 
\textbf{Sadness} & $.35$ & $.42$ & $-.141$** & $-.172$**\\ 
\textbf{Fear} & $.43$ & $.43$ & $-.121$** & $-.117$**\\
\textbf{Anger} & $.81$ & $.66$ & $-.217$** & $-.295$** \\
\textbf{Surprise} & $.03$ & $.08$ & $-.011$ & $-.016$ \\ \hline

\multicolumn{5}{|c|}{\textbf{Negotiation Outcomes}} \\
\textbf{Satisfaction} & $4.17$ & $1.03$ & $1$ & $.702$\\ 
\textbf{Likeness}& $4.11$ & $1.12$ & $.702$ &  $1$ \\ \hline
\end{tabular}}
\label{tab:variables-stats}
\end{center}
\end{table}

\section{Analysis of Emotion Variables}
\label{sec:analysis}
Before presenting our results for predicting participant satisfaction and likeness, we first validate \textit{T5-Emotion} measures by understanding how they relate to \textit{Emoticon} and \textit{LIWC}.

\subsection{Correlation among emotion dimensions} Table \ref{tab:correlations-among-emotion} presents the correlations of \textit{T5-Emotion} measures among each other and with \textit{Emoticons} and \textit{LIWC} dimensions. For this analysis, we only focus on the dimensions that appear in all three models: \textit{Joy}, \textit{Sadness}, and \textit{Anger}. For \textit{LIWC}, we incorporate the \textit{Positive} dimension. In general, we find significant positive correlations of the emotion dimensions with their own counterparts across the three models, which aligns well with our expectations and validates the model predictions on the CaSiNo dataset. We observe that although significant, most correlation values are weak to moderate, indicating the fundamental differences in what is captured by these three models. We also note that \textit{Anger} shows significant positive correlations with \textit{Sadness} as well. Since these measures are based on the confidence scores of the model, this suggests that the model might get confused between these two negative emotion dimensions.

\subsection{Lexical correlates of emotion} To gain more insights, we perform additional qualitative analysis by generating lexical correlates for the emotion measures. We assigned an emotion label to each utterance for all three methods. For \textit{Emoticons} and \textit{LIWC}, this was based on the label that is reflected in the majority in the utterance. For \textit{T5-Emotion} dimensions, we simply used the final predictions from the model. We next compute the log-odds ratio, informative Dirichlet prior~\cite{monroe2008fightin} of all the tokens for each emotion dimension relative to all other dimensions. This allows us to study the associations between the tokens and each emotion dimension. We present the top $5$ words in each category in Table \ref{tab:liwc-log-odds} along with scaled Z-scores. We observe a number of highly significant associations across all categories that align with our intuition such as words like `awesome', `great', `thank' for \textit{Joy} and \textit{Positive Emotions} categories. We found these words to be common in the beginning and end of the negotiation where participants engage in small talk, resulting in such utterances being classified as \textit{Joy} or \textit{Positive}. Further, the associations are higher for \textit{LIWC} and \textit{T5-Emotion} which, as we observe, are able to recognize many other emotive utterances. Although significant, we observe that some words are less meaningful such as in the case of \textit{Anger}. This suggests that some emotion dimensions may require more utterance-level context for better interpretation.

\subsection{Sample Predictions}
For a more contextual analysis, we explicitly observe the sample predictions for \textit{T5-Emotion}. We still focus on the three dimensions that are the most common and are present in all the three emotion methods. Table \ref{tab:t5-sample-predictions} presents a few of the most confident predictions by the model. To analyze the utterances that are specifically captured by the deep learning model, we only look at the utterances that are \textit{undetected} by the other two methods. As the table depicts, the model is able to capture contextual emotion, beyond just the lexical or emoticon usage. Based on these observations, we expected the affective attributes extracted using the deep learning model to be better predictors than \textit{Emoticons} and \textit{LIWC} measures.

\section{Results}
\label{sec:results}
We now present our main results.

\subsection{Correlation with outcomes}
Table \ref{tab:variables-stats} summarizes the correlations of all the continuous measures defined in Section \ref{sec:dataset} with the negotiation outcomes. Overall, we observe a number of significant trends across individual differences and emotion variables. We find that satisfaction and partner likeness both improve with age ($r$=$0.110$, $p<0.01$). The outcomes are also positively correlated with Agreebleness and Emotional Stability ($ps<0.01$). 
We observe that positive emotions tend to relate positively with both the outcomes, showing significant trends for likeness in the case of \textit{Emoticon} usage ($r$=$0.075$, $p<0.01$) and \textit{LIWC} ($r$=$0.061$, $p<0.01$). We do not observe similar results for \textit{T5-Emotion}. This may be due to the inherent bias that the deep learning model shows towards \textit{Joy} label, reducing its precision. However, trends show significant negative correlations for both \textit{Sadness} and \textit{Anger} that are usually expressed when the negotiation is not going favorably. This is naturally associated with lower satisfaction of the participants and liking for their negotiation partners.

Among the discrete variables, we find that females are significantly more satisfied ($t$=$3.6$, $p<0.001$) and significantly more like their partners ($t$=$4.1$, $p<0.001$) than males. We do not observe any significant associations with SVO and Ethnicity, based on t-test and one-way ANOVA respectively.

\subsection{Regression Analysis}
We now discuss the results of our regression analysis that answers whether emotion variables extracted from the negotiation dialogue collectively explain more variance in satisfaction and likeness, above and beyond the individual difference variables. To achieve this, we perform regression with three steps where each subsequent step incorporates the following set of variables: 1) Individual difference, 2) Affect variables of the participant, and 3) Affect variables for the negotiation partner. The affect variables come from one of the emotion methods defined in Section \ref{sec:dataset}, which also helps to compare their performance. As discussed earlier, these variables only encompass the information about the negotiation that would be entirely \textit{visible} to an automatic negotiation agent, either explicitly from the negotiation dialogue, or implicitly by inferring the individual difference attributes from past online behaviors of the users on social media platforms. Hence, our insights from such an analysis can practically aid in designing sophisticated negotiation agents that incorporate user satisfaction and their perception of the agent itself in their modelling.

Table \ref{tab:regression-satisfaction} and \ref{tab:regression-likeness} summarize the results for satisfaction and likeness respectively. Overall, we observe that the predictions are highly significant with just using the individual difference variables. However, incorporating the affective attributes of the participant and the negotiation partner explains significantly more variance.

For satisfaction, the individual difference variables account for minimal yet significant variance (F($14$, $1997$)=$3.46$, $p<.001$, R$^2$=$.024$). Adding the participant's affect variables based on the \textit{T5-Emotion} model in the second step, helps to account for a much higher variance (F($20$, $1991$)=$10.41$, $p<.001$, R$^2$=$.095$), such that this increase in the proportion is itself highly significant ($\Delta$F($6$, $1991$)=$26.02$, $p<.001$, $\Delta$R$^2$=$.071$). Yet, further variance is explained when the partner's affect variables are incorporated (F($26$, $1985$)=$10.88$, $p<.001$, R$^2$=$.125$: $\Delta$F($6$, $1985$)=$11.38$, $p<.001$, $\Delta$R$^2$=$.030$).

Similar trends can be observed for likeness as well. Individual difference variables alone  account for a significant proportion (F($14$, $1997$)=$6.05$, $p<.001$, R$^2$=$.041$) but adding the participant's \textit{T5-Emotion} attributes further improves the prediction performance (F($20$, $1991$)=$18.16$, $p<.001$, R$^2$=$.154$) with a significant increase ($\Delta$F($6$, $1991$)=$44.58$, $p<.001$, $\Delta$R$^2$=$.114$). Finally, when the partner's attributes are also incorporated as predictors, this again shows significant improvements in the explained variance (F($26$, $1985$)=$19.07$, $p<.001$, R$^2$=$.200$: $\Delta$F($6$, $1985$)=$18.83$, $p<.001$, $\Delta$R$^2$=$.046$).

Lastly, we note that in line with our linguistic analysis, the ability for \textit{T5-Emotion} to capture more contextual emotive utterances helps to account for more variance, as compared to \textit{Emoticons} and \textit{LIWC} independently. In our experiments, we also observed that even when \textit{Emoticons} and \textit{LIWC} are combined, \textit{T5-Emotion} still achieves a much better performance.


\begin{table}[t!]
\caption{Regression results for predicting Satisfaction. ** denotes $p<0.01$. *** denotes $p<0.001$.}
\begin{center}
\scalebox{0.8}{
\begin{tabular}{|l|c|c|c|c|c|}
\hline
\textbf{Variables} & $\mathbf{R^2}$ & \textbf{df} & \textbf{F} & \textbf{$\mathbf{R^2}$ Change} & \textbf{F Change} \\ \hline
\multicolumn{6}{|c|}{\textbf{Affect Variables: Emoticons}} \\
\textbf{Individual Difference} & $.024$ & ($14$, $1997$) & $3.46$*** & $-$ & $-$\\
\textbf{+Participant Affect} & $.051$ & ($18$, $1993$) & $5.92$*** & $.027$ & $14.20$*** \\
\textbf{+Partner Affect} & $.063$ & ($22$, $1989$) & $6.04$*** & $.012$ & $6.34$***\\ \hline
\multicolumn{6}{|c|}{\textbf{Affect Variables: LIWC}} \\
\textbf{Individual Difference} & $.024$ & ($14$, $1997$) & $3.46$*** & $-$ & $-$\\
\textbf{+Participant Affect} & $.032$ & ($18$, $1993$) & $3.61$*** & $.008$ & $4.07$**\\
\textbf{+Partner Affect} & $.051$ & ($22$, $1989$) & $4.85$*** & $.019$ & $10.14$***\\ \hline
\multicolumn{6}{|c|}{\textbf{Affect Variables: T5-Emotion}} \\
\textbf{Individual Difference} & $.024$ & ($14$, $1997$) & $3.46$*** & $-$ & $-$\\
\textbf{+Participant Affect} & $.095$ & ($20$, $1991$) & $10.41$*** & $.071$ & $26.02$*** \\
\textbf{+Partner Affect} & $.125$ & ($26$, $1985$) & $10.88$*** & $.030$ & $11.38$*** \\ \hline
\end{tabular}}
\label{tab:regression-satisfaction}
\end{center}
\end{table}

\begin{table}[t!]
\caption{Regression results for predicting Likeness. *** denotes $p<0.001$.}
\begin{center}
\scalebox{0.8}{
\begin{tabular}{|l|c|c|c|c|c|}
\hline
\textbf{Variables} & $\mathbf{R^2}$ & \textbf{df} & \textbf{F} & \textbf{$\mathbf{R^2}$ Change} & \textbf{F Change} \\ \hline
\multicolumn{6}{|c|}{\textbf{Affect Variables: Emoticons}} \\
\textbf{Individual Difference} & $.041$ & ($14$, $1997$) & $6.05$*** & $-$ & $-$\\
\textbf{+Participant Affect} & $.080$ & ($18$, $1993$) & $9.68$*** & $.040$ & $21.50$*** \\
\textbf{+Partner Affect} & $.097$ & ($22$, $1989$) & $9.70$*** & $.017$ & $9.12$***\\ \hline
\multicolumn{6}{|c|}{\textbf{Affect Variables: LIWC}} \\
\textbf{Individual Difference} & $.041$ & ($14$, $1997$) & $6.05$*** & $-$ & $-$\\
\textbf{+Participant Affect} & $.055$ & ($18$, $1993$) & $6.39$*** & $.014$ & $7.32$***\\
\textbf{+Partner Affect} & $.072$ & ($22$, $1989$) & $7.06$*** & $.018$ & $9.56$***\\ \hline
\multicolumn{6}{|c|}{\textbf{Affect Variables: T5-Emotion}} \\
\textbf{Individual Difference} & $.041$ & ($14$, $1997$) & $6.05$*** & $-$ & $-$\\
\textbf{+Participant Affect} & $.154$ & ($20$, $1991$) & $18.16$*** & $.114$ & $44.58$*** \\
\textbf{+Partner Affect} & $.200$ & ($26$, $1985$) & $19.07$*** & $.046$ & $18.83$*** \\ \hline
\end{tabular}}
\label{tab:regression-likeness}
\end{center}
\end{table}

\section{Discussion and Conclusions}
\label{sec:discussion}
Our aim was to empirically investigate the extent to which the affect variables extracted from the negotiation itself are helpful for prediction of two important metrics for negotiation agents – participant's outcome satisfaction and liking for the partner. We presented an extensive analysis based on a large-scale dataset of human-human negotiation dialogues, grounded in a realistic camping scenario. We devised three degrees of emotion dimensions, from leveraging emoticon usage similar to prior work in menu-driven systems, to going beyond and extracting emotion expression directly from the textual chat utterances. Our results show that such affect variables explain variance in these subjective outcomes, above and beyond the impact of individual difference variables that are available before the negotiation begins\footnote{We also tried controlling for the objective negotiation performance of the partner in the regression analysis. We still observed similar significant trends for both satisfaction and likeness.}.

One might expect that the affect variables would help merely because they are manifestations of the individual difference such as the social value orientation, agreeableness, or gender. If this was indeed the case, there would have been no need to incorporate such affective variables, but merely the individual difference variables would have sufficed. However, to the extent to which the affect variables help in predictions, above and beyond these individual differences, this suggests that there is utility in extracting these emotion dimensions explicitly for improved agent's performance, being especially well-suited for more subjective outcomes in a negotiation.

We further note that the individual difference variables that we use are based on self-identification and standard personality tests from the psychology literature. This makes these variables more reliable than if they were to be implicitly inferred from past social behaviors of the users, as attempted in a number of previous works analysing interactions on social media platforms~\cite{dong2014inferring, ortigosa2011inferring, adali2014predicting}. If affective factors show significant improvements above the individual differences measured with minimal error, this further attests to their utility when the demographics and personality information is inferred, and thus, less reliable.

Since negotiations and other mixed-motive situations can be  fraught with emotional decisions, our findings suggest that it would be useful for designers of agents that negotiate with humans to be \textit{armed} with algorithms for understanding the unfolding emotions displayed by users in terms of their emoticon use and natural language. Our work suggests that demographic and personality attributes inferred from past social interactions (such as on social media) are not sufficient, and that providing information to negotiating agents from affective channels tracked during the negotiation itself is important for developing agents that can predict or understand if their human counterpart is satisfied and likes them. Ultimately, outcomes like satisfaction and likeness will be essential for such agents to cultivate if they, like human negotiators, hope to successfully negotiate with that same partner in the future~\cite{aydougan2020challenges}. 


In light of the theme for this year's ACII on \textit{Ethical Affective Computing}, we briefly discuss the ethical considerations around the development of automated negotiation agents. There has been a significant amount of work on ethics in negotiation literature. The acts of emotion manipulation, deception, bias, and misinterpretation have been central concerns~\cite{lewicki2016essentials}. It is plausible that these behaviors also get reflected in the negotiation agents that are grounded in the insights gained from human-human negotiation datasets such as CaSiNo. To mitigate the impact of these ethical concerns when automated negotiation systems are deployed to the end users, we lay out a few recommendations. We believe that transparency is the key. Being upfront and open about the identity, capabilities, and any known misbehaviors of the agent would prevent any misinterpretation and would set the right expectations for the end users. Further, we note that our work encourages the agent to adapt its behavior by incorporating affective attributes along with user demography and personality. If the deployed agent is indeed adaptive towards these traits, we recommend a regular monitoring cycle for unexpected behaviors of the agent, preventing against any offensive or discriminative actions.
\bibliographystyle{IEEEtran}
\bibliography{mybibliography}

\end{document}